\documentclass{sig-alternate-05-2015}
\usepackage{cite}

\usepackage{subcaption}

\usepackage{array,multirow,makecell}
\newcolumntype{C}[1]{>{\centering\arraybackslash }b{#1}}

\usepackage{url}

\usepackage{comment}
\excludecomment{comment} 

\begin{document}

\conferenceinfo{HIP3ES, 2016}{January 18--20, 2016, Prague, Czech Republic}

%
\CopyrightYear{2016}
\doi{000.0000}
\isbn{000-0-0000-0000-0}
%

\title{Fast Power and Energy Efficiency Analysis of FPGA-based Wireless Base-band Processing}
%
%
%
%
%

\numberofauthors{1} 
%
\author{
%
%
\alignauthor
Jordane Lorandel, Jean-Christophe Pr\'evotet and Maryline H\'elard\\ 
       \affaddr{INSA of Rennes - Institute for Electronics and Telecommunications of Rennes}\\
       \affaddr{22 avenue des buttes de coesmes,Rennes, France}\\
       \email{jordane.lorandel@insa-rennes.fr}
}

\maketitle
\begin{abstract}
Nowadays, demands for high performance keep on increasing in the wireless communication domain. This leads to a consistent rise of the complexity and designing such systems has become a challenging task. In this context, energy efficiency is considered as a key topic, especially for embedded systems in which design space is often very constrained. 
In this paper, a fast and accurate power estimation approach for FPGA-based hardware systems is applied to a typical wireless communication system. It aims at providing power estimates of complete systems prior to their implementations. This is made possible by using a dedicated library of high-level models that are representative of hardware IPs. Based on high-level simulations, design space exploration is made a lot faster and easier. The definition of a scenario and the monitoring of IP's time-activities facilitate the comparison of several domain-specific systems. The proposed approach and its benefits are demonstrated through a typical use case in the wireless communication domain.
\end{abstract}


\\
\\
\\



%


\section{Introduction}  \label{sec:intro}
Today, mobile networks continuously evolve to deal with several issues such as the data traffic growth, the increasing number of user equipments, the multiplication of standards, etc. Authors in \cite{Cisco15} show that almost half a billion (497 million) mobile devices and connections were introduced in 2014. It is also foreseen that global mobile data should have a compound annual growth rate of 57\% from 2014 to 2019. This clearly underlines the need of perpetual evolution of all the actors dealing with mobile networks.
The processing capabilities of such systems will also drastically increase due to the perpetual demand for higher functionality and performance.

For the last decades, power has become one of the most important issue and has been the subject of numerous researches in the wireless communication field \cite{Earth15,GreenTouch15,5Green15}. Achieving a high level of performance in terms of throughput or low latency, still constitutes a primary objective. 
Design complexity and the time that is required to develop such systems drastically increase. It is of particular importance for designers to make early design choices to respect their power budget. Consequently, designers have to estimate power as soon as possible in the design process. 
Moreover, decisions that are taken very late in the design flow could lead to important additional costs and generally impose to rerun the design process from scratch.

Today, Field Programmable Gate Array (FPGA) devices are widely used as solutions to implement complex designs such as wireless base-band processors. As compared to their ASIC (Application Specific Integrated Circuit) counterparts, which can achieve higher performance at a price of a long and expensive design development, FPGAs can be used for fast and low cost ASIC prototyping or as hardware accelerators for real-time applications. FPGA-based systems can be made of IP (Intellectual Property) which are hardware cores that enhance design reuse and speed-up development time. 
FPGAs can offer more flexibility than ASICs to the detriment of a higher silicon area, a decrease in performance and a higher power consumption \cite{Kuon2007}. 


\section{Positioning of the approach} 

From the FPGA description, it seems obvious that estimating power very fast in the design process is a primary objective. However, power estimation usually requires simulations that can be very time consuming depending on the level of abstraction. The higher the level is, the faster power results are obtained. Alternatively, although a good accuracy can be achieved at low-level, simulation times are often prohibitive.
Since accuracy is decisive during design space exploration, designers generally consider a small set of accurate configuration patterns. They do not have enough time to test an exhaustive set of examples that could lead to a better design solution.

Another consideration is that there is a lack of methodologies and approaches that enable domain-specific systems comparison in terms of both power and performance in an efficient way. Indeed, system performance are usually evaluated at high-level whereas accurate power information is only available at lower levels when a hardware target is considered. In addition, several teams of designers are usually required from the system's specification to the real hardware implementation. As consequence, there is still a real gap to bridge between high-level and low-level that can be error prone.

Wireless communication systems are almost made of Power Amplifiers (PA), RF stages and base-band processing (BB). In such system, PA enable data transmission over the air. In fact, the power allocated for data transmission is usually considered as the most significant contributor in a wireless communication system. As a consequence, power consumptions that are related to BB and RF are usually neglected. 
However, it has been demonstrated that all power consumption sources have to be taken into account in a wireless system, especially the power consumption related to the base-band processing when low transmission powers are involved \cite{EarthD2_3}. As an example, the power consumption of BB may represent around half of the total power consumption for a base station of a femto/home cell-environment in the LTE context. Such conclusions advocate an efficient evaluation of the BB power consumption. 

The purpose of this paper is to present a power estimation approach for FPGA-based wireless communication systems. The detailed description of the methodology is not presented in this paper. 
We rather demonstrate its benefits through a typical wireless communication system. The methodology deliberately focus on hardware designs without taking any software considerations into account. A key contribution consists in taking into account IP-time activities that directly evolve according to the application behaviour. We also demonstrate that power estimates can be thus refined, providing more accurate results than classic approaches. We have also noticed an important speed-up factor. 

This paper is organized as follows: Section \ref{sect_RW} presents the related works on high level power estimation. Section \ref{sect_Meth} promotes the main contributions of the proposed methodology. Section \ref{sect_usecase} provides the results that have been obtained by applying our methodology to a MISO-OFDM 2x1 system in the Long-Term Evolution context. Finally, we conclude and discuss about prospects in section \ref{sect_conclu}.

\section{Related Works}\label{sect_RW}
\begin{figure}[!t]
\centering
\includegraphics[scale=0.6,trim=0.5cm 0cm 0cm 0cm, clip=true]{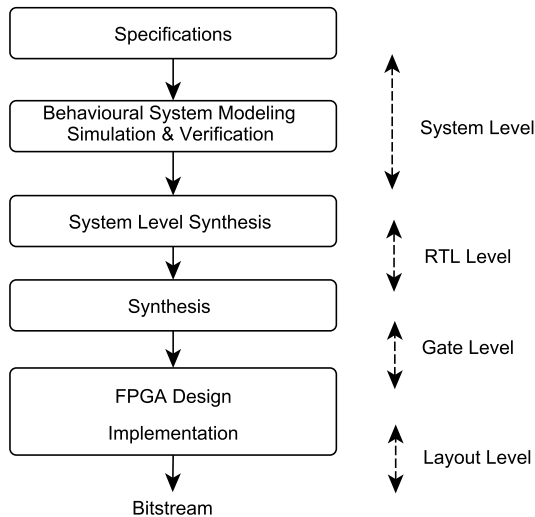}  
\caption{Representation of the different abstraction levels from specification to FPGA bitstream generation}
\label{Abstraction_Levels_label}
\end{figure}

Estimating power in FPGA devices has been the subject of various research studies. It has been noticed that power estimation techniques may be applied at different level of abstraction, which corresponds to the degree of details to describe a system. As shown in Fig. \ref{Abstraction_Levels_label}, the lowest level of abstraction is the layout-level which corresponds to a bitstream model in a FPGA design flow.
At layout, gate and RTL levels, power estimation can be very accurate due to the prominence of physical and hardware details. However, power estimation can be really costly in time. It also forces designers to take decisions very late in the design flow, which may lead to expensive redesign costs if constraints are not met. Due to the growing complexity of current systems, it is obvious that low-level approaches are not suitable for fast power estimations.
The highest abstraction level is the system-level, in which the functionality of a system is modelled using specific languages and dedicated tools. At this level, implementation details are completely hidden or not available. Although the simulation time is highly reduced, power estimation accuracy is usually lower than using low-level approaches. Nevertheless, designers may explore design architectures very fast and make early decisive choices. This is especially true when the primary objective is the design of low-power systems \cite{ITRS}.
For this reason, we deliberately focus our study on system-level power estimation tools and methodologies.

General power values can be used to estimate the power consumption of wireless communication systems \cite{Desset12, Deruyck12, Arnold10}. Average power consumption values are associated with different elements in the system. Using such power models, accuracy is generally not the the main purpose and only global performance trends are provided.

In the wireless communication domain, designers use to work with high-level modelling tools such as Matlab/Simulink from Mathworks \cite{Math15}. Some other academic and industrial tools are also utilized to a lesser extent, in the signal processing and in the wireless communication domain \cite{Pto15,Lab15,System15, Vista_Arch_Mentor}. Basically, such tools do not support power estimation for FPGA-based systems. In fact, an additional tool is required to complete the FPGA design flow before obtaining any information related to the power consumption of their algorithms. As an example, Xilinx System Generator \cite{SysGen15} enables to realize the FPGA design implementation directly from a Matlab/Simulink description. Accurate hardware details from lower levels can thus be obtained. However, design space exploration is really limited due to the numerous iterations that are required for FPGA design implementation. A similar approach is followed in a Matlab/Simulink add-on tool that is described in \cite{PyGen06}. With this tool, detailed low-level information based on the synthesis of FPGA designs, is forwarded to higher levels using System Generator.

System modelling is an interesting approach to deal with the increasing complexity of digital systems. Programming languages such as C/C++ or high-level description languages such as SystemC \cite{Accellera15}, are more and more used in the community. The latter enables both software and hardware description and supports several degrees of design refinement that are well-suited for embedded systems. Moreover, SystemC-Transaction-Level Modelling (TLM) has been successfully standardized by the Open SystemC Initiative (OSCI) \cite{montoreano2007transaction}.

However, estimating power in systems that have been modelled using such languages is basically not supported, as for Matlab. Indeed, designers have to first enrich the language with specific information before performing any estimation.  
Based on these considerations, several works have contributed to this problem in order to make power estimation possible at system-level. Macro-models can be used to directly link power to the signal statistics of the circuits \cite{Dama07}. 
Power consumption can also be estimated through the definition of power state machines (PSM) that model each core of the system \cite{Hsu11}. Power models of each IP are integrated into SystemC models to enable power high-level power estimation. In this work, low-level simulations are first realized to build accurate models and simulations are then performed to evaluate the overall power consumption. Note that IP execution modes are identified using key signals during these simulations. 

Functional power models for IP cores can be built according to the Functional-Level Power Analysis methodology (FLPA) \cite{Elleouet06}.
When combining with a SystemC-TLM approach, such power models can also be used to estimate the power consumption of MPSoC targets \cite{Rethinagiri11}.

FPGA vendors have also developed early power estimation tools called spreadsheets \cite{XPE12,Alt14}. Such sheets represent another way of estimating the average power consumption of a system based on FPGA. Using accurate FPGA hardware details, they provide early power estimates according to user specifications, prior to any implementation. In fact, analytical formulas aim at quantifying power consumption based on user-defined values such as the number of resources, clock frequency, signal activity, voltage, etc. Also, spreadsheet power estimates can be refined all along the design flow. Nevertheless, obtaining accurate power estimations is generally a long and difficult task, especially when large and complex systems are considered.


According to previous studies, numerous techniques are available at different levels of abstraction. At system level, if low-level details are not available, such lack of information may lead to a poor power estimation accuracy. The same conclusion can be drawn using general analytical models when power consumption is estimated using average power consumption values. Such results are very interesting in practice but they may not be appropriated to the design of complex systems. Moreover, such results are not easily scalable since they are usually devoted to a specific and dedicated hardware. Furthermore, functionality is rarely validated with power estimation, which is is an important issue if designers want to evaluate the power and performance trade-off at the same time. In this way, SystemC-TLM approach seems to be a promising technique to deal with these issues. 
From our knowledge, there is a lack of system-level tools that enable to perform an efficient comparison between several configurations of FPGA-based systems in terms of power consumption and performance, especially in the wireless communication domain. Although, cores power modelling is interesting, it usually does not consider performance and only focus partially on the system. 
 
From these considerations, we propose an approach that is dedicated to FPGA-based hardware wireless communication designs. It aims at addressing several points: First, the proposed approach strives to provide a fast feedback of accurate power estimations from low-level components to the system level. To this purpose, the approach is based on the development of a dedicated library of cores and models that will be used by designers. Second, the concept of scenario is introduced to efficiently compare several applications. Third, the simulation time is significantly reduced, while obtaining a satisfactory level of accuracy that is similar to gate-level results. 

\section{Power Estimation Approach}\label{sect_Meth}
The proposed approach is dedicated to hardware system without any software consideration. Moreover, it is assumed that wireless communication systems can be entirely described using a set of interconnected hardware IP cores that constitute a data-flow architecture. Such IPs are usually dedicated to a specific function e.g. Fast Fourier Transform (FFT), channel encoders, modulators, etc. 
First, the entry point of the methodology is the innovative definition of a scenario. This term has already been used in \cite{gheorghita2009system, zompakis2013scenario} but with another sense. In our approach, this concept refers to a set of parameters which are common to several applications in the same domain. A scenario is composed of system and technological parameters which have both an effect on power and/or performance. 
Using this concept, common features of several applications are clearly identified and other features can be evaluated. In this way, a comparison of several systems can be performed in an efficient way based on the observation of the impact of each parameter over the power and performance trade-off. This is of particular interest regarding the complexity and the high number of user-defined parameters from different levels. 

From this definition, each application corresponds to an instance of a scenario, which also means that a scenario can be seen as a meta-model of an application. As the proposed example given in Table \ref{Table_example_Scenario_label}, an application refers to a fully parametrized scenario. In the wireless communication domain, this may refer to a modulation type, a specific coding scheme, a frequency bandwidth, and other technological parameters such as the clock frequency, a FPGA target, the data quantization (number of bits to represent a modulated symbol), etc.

\begin{table}[!t]
\caption{Example of a scenario with two applications}
\label{Table_example_Scenario_label}
\centering
\begin{tabular}{|c|c|c|}
\hline\hline
Scenario Parameters &  Application 1 &  Application 2   \\
\hline\hline
 Channel Coding Rate & \multicolumn{2}{c|}{1/3}  \\
\hline
 QAM Modulation &  QPSK & 16-QAM \\
\hline
 Frequency Bandwidth &  5MHz  &  10MHz  \\
\hline
 Data quantization &  10 bits &  14 bits \\
\hline
  Frame Type &  \multicolumn{2}{c|}{  1 OFDM symbol pilot } \\	
 & \multicolumn{2}{c|}{ every 10 OFDM symbols of data} \\
\hline
 FPGA & \multicolumn{2}{c|}{ Xilinx Virtex-6 LX240T} \\
\hline
 Clock Frequency & \multicolumn{2}{c|}{ 100 MHz} \\
\hline\hline
\end{tabular}
\end{table}

The proposed methodology is realized in two stages: an IP characterization and modelling phase using SystemC, followed by a global system simulation.
Note that the first stage is only realized if IPs are not in the library. If a specific IP is not in the library, the first stage of the methodology consists in enriching the library with the associated power values and models. To this purpose, hardware IPs are fully characterized in terms of power and behaviour, at both low and high levels. A gate-level power estimation tool from Xilinx called XPower Analyzer (XPA) has been used to estimate the average power consumption. To obtain accurate power estimation, the implementation design flow has been realized for each IP. A post-place and route VHDL simulation model, including timing properties, has been generated under the Xilinx design software environment (ISE v14.4). This model has then be simulated using the Modelsim SE-64 10.1c \cite{Mentor15} tool in order to record the internal activity of all elements that constitute the IPs. Based on specific test-benches, two simulations have been performed. The first one was performed when the IP is active and the other one when the IP is idle. Finally, power has been estimated using XPA according to the activity file and additional implementation files i.e. constraint and design netlist files. Note that XPA delivers a complete report on the average power used by the different elements of the FPGA i.e. clock, logic, signals, memories, DSP blocks, etc.



The characterisation process using XPA and timing simulations have to be re-run several times for each set of parameters. The characterisation process is effective when all power estimation results have been obtained for every IP of the system. This first stage is quite tedious but has been relieved by the use of automated scripts that also spare time and reduce the number of errors.

After performing IP power evaluations for each configuration, information is added to the corresponding SystemC description. Each SystemC model is built with respect to a specific implementation model that is composed of a control part and a data path. 
In addition, key signals of the corresponding hardware IP are represented at high level to identify if the IP is active or not and thus determine which power values has to be considered. 
"Clock enable" and "read/write" input signals are common examples of key signals that are shared and used by hardware IPs.
Moreover, some FPGA vendors provide bit-accurate C-models of their hardware IP. They can be easily integrated with SystemC in order to accurately model the functionality of the hardware IP.  
Each high-level model is then stored in a dedicated library for further reuse.

\begin{figure}[!t]
 \centering
 \includegraphics[scale=0.47,trim=0cm 25cm 0cm 0cm, clip=true] {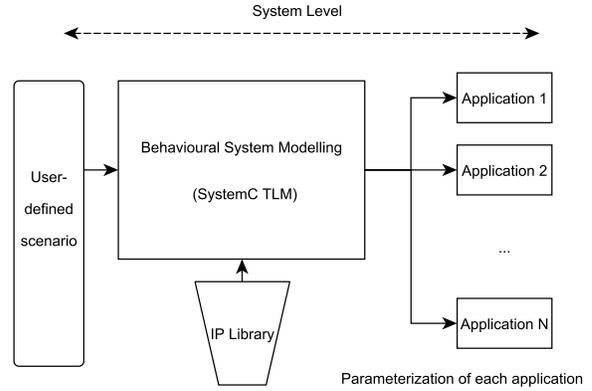} 
 \caption{Second step of the methodology based on the scenario}
 \label{Scenario_approach_label}
 \end{figure}

\begin{figure}[!t]
 \centering
 \includegraphics[scale=0.35,trim=0.5cm 0cm 0cm 0cm, clip=true] {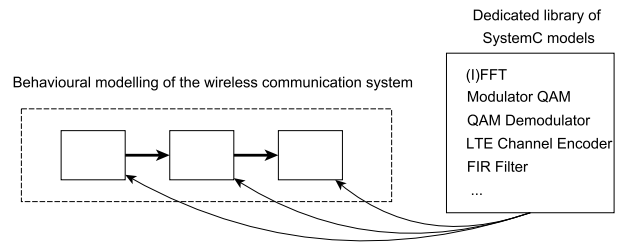} 
 \caption{Behavioural modelling and simulation using SystemC}
 \label{Fig_2nd_step_method_label}
 \end{figure}

The second stage of the methodology is described in both Fig. \ref{Scenario_approach_label} and Fig. \ref{Fig_2nd_step_method_label}. This stage is the entry point for designers. Their systems are built by connecting the SystemC models that have been developed and stored in the library during the first stage. Once the SystemC model of the entire system is described, users can define their applications i.e. instances of scenario, by setting up system and technological parameters. System functionality can also be validated using the SystemC simulation kernel. 

From the system model, simulations are performed and time-activity coefficients of all IP models are obtained regarding the key signals' evolution. Such coefficients represent the percentage of the simulation time during IPs are active. Indeed, it is of particular importance to take into account the temporal activity of each IP, that highly depends on the application. In fact, for a given circuit, each application behaviour may have a significant impact on the final power estimation results. 
Finally, the power consumption of the entire system is estimated by determining the power contribution of each IP that builds the system and their corresponding time-activity coefficients.

To demonstrate the benefits of the proposed approach, a fully-compliant Long-Term Evolution (LTE) wireless base-band transmitter has been developed. 
\section{Use Cases}\label{sect_usecase}
 
Long-Term Evolution (LTE) is the fourth generation (4G) of radio technology for mobile wireless communications that has been standardized by the 3rd Generation Partnership Project (3GPP) cooperation. LTE has multiple objectives such as the latency reduction, throughputs improvements or the users management.






In order to achieve high performance, the LTE physical layer combines several technologies such as Orthogonal Frequency Division Multiple Access (OFDMA) for downlink (DL) and Single Carrier Frequency Division Multiple Access (SC-FDMA) for Uplink (UL). It enables to achieve a throughput up to 100 Mbs in DL and up to 50 Mbs in UL. These values can also be improved when considering the last specifications of LTE-A i.e. carrier aggregation and other improvements. 

The main OFDM parameters in LTE \cite{LTE_norm} are given in Table \ref{Table_OFDM_Param_label}. As described in this table, LTE uses scalable bandwidths from 1.4 MHz up to 20MHz. It also enables the use of Multiple Inputs Multiple Outputs (MIMO) schemes to improve systems performance. The MIMO strategy increases complexity, while improving spectral efficiency (when combined with OFDM), data throughput, and robustness to interferences. In the LTE standard, several MIMO operating modes are available. Transmit diversity, open-loop and closed-loop spatial multiplexing for a single user (up to 4x4 in Release 9 for DL and up to 8x8 in Release 10) are available techniques for DL \cite{Li10}.

The LTE frame has a 10 ms duration. It carries and exchanges specific data based on different physical channels and signals. Two structures of frame are also standardized in LTE. Here, we focus on the type 1 structure that supports Frequency Division Duplex (FDD) whereas type 2 deals with Time Division Duplex (TDD). We also focus on the Physical Downlink Shared CHannel (PDSCH) and its associated processing that is dedicated to user data. Other channels are not studied in this paper.

\begin{table}[!t] 
\renewcommand{\arraystretch}{1.3}
\centering 
\caption{ Main Downlink OFDM parameters in LTE }
\begin{tabular}{|p{2cm}|c|c|c|c|c|c|} 
\hline\hline 
 Spectral Bandwidth (MHz) & 1.4 & 3& 5& 10 & 15& 20 \\
\hline
  Sub-carrier spacing (kHz)  & \multicolumn{6}{c|}{ 15} \\   
\hline
 Used sub-carriers & 72 & 180 & 300 & 600 & 900& 1200\\  
\hline
 Used Resource Blocks  & 6 & 12 & 25 & 50 & 75 & 100 \\
\hline
 (I)FFT Size  & 128 & 256& 512& 1024& 1536 & 2048 \\  
 \hline
  OFDM symbol Length  & \multicolumn{6}{c|}{66.67$\mu s$ (without Cyclic Prefix)} \\
 \hline 
 Cyclic Prefix & \multicolumn{6}{c|}{normal: 5.21$\mu s$ (1st symbol) then 4.67$\mu s$}\\
  Length & \multicolumn{6}{c|}{extended: 16.67$\mu s$}  \\
 \hline \hline
\end{tabular} 
\label{Table_OFDM_Param_label} 
\end{table}

 \begin{figure*}[!t]
 \centering
 \includegraphics[scale=0.45,trim=0.0cm 0cm 0cm 0cm, clip=true]{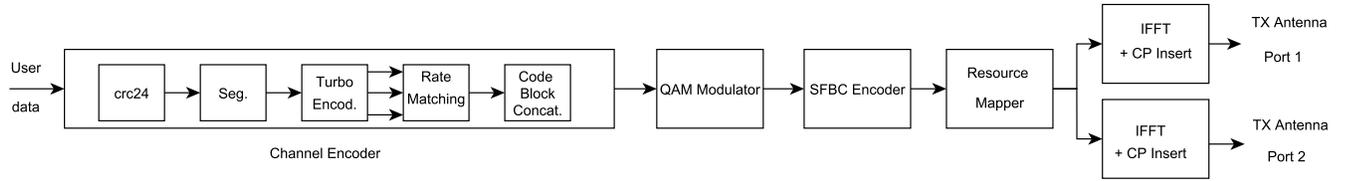}  
 \caption{LTE TX Downlink processing for PDSCH channel in MISO2x1 configuration with Alamouti encoding}
 \label{LTE_MISO2x1_label}
 \end{figure*}

In Fig. \ref{LTE_MISO2x1_label}, the Downlink PDSCH transmitter processing, in a MISO 2x1 OFDM configuration, is shown. This design has been entirely developed using the VHDL Hardware description language in order to efficiently compare the results of our approach with a real system. 


During the development of this system, we have made the assumption that any resources of a LTE frame that are not dedicated to PDSCH channel are deliberately set to 0. 
A first scenario has been defined in order to compare the different applications that are summarized in Table \ref{lte_scenario_and_app_label}. It can be noticed that IFFT sizes are different for the four studied applications. This parameter has an impact on both power consumption and system performance. 

\begin{table}[!t]
\caption{Defined scenario and applications}
\label{lte_scenario_and_app_label}
\centering
\begin{tabular}{|c|c|c|c|c|}
\hline\hline
\multirow{2}{*}{Scenario} & Appli.  &  Appli. &  Appli.&  Appli. \\
&  1 &  2 &  3&  4 \\
\hline\hline
 \multirow{2}{*}{ Channel Coding } & \multicolumn{4}{c|}{ Rate=1/3}\\ 
 &  \multicolumn{4}{c|}{ Code block size = 1024} \\
\hline
QAM Modulation & \multicolumn{4}{c|}{QPSK} \\
\hline
 (I)FFT Size &  256  &  512 &  1024 &  2048  \\
\hline
 TX antenna & \multicolumn{4}{c|}{2}  \\
\hline
 Data quantization & \multicolumn{4}{c|}{14 bits} \\
\hline
  FPGA Type & \multicolumn{4}{c|}{ Xilinx Virtex-6 LX240T} \\
\hline
 Clock Frequency & \multicolumn{4}{c|}{ 50 MHz} \\
\hline
 Simulation time & \multicolumn{4}{c|}{ Generation of 5 LTE sub-frames} \\
\hline\hline
\end{tabular}
\end{table}

\subsection{Power estimation}

The two stages of the approach have been applied to the proposed system. Each IP that composes the hardware system has been characterized independently. Moreover, SystemC models were built according to the proposed methodology. System-level simulations have been performed for the four applications and corresponding IP time-activities have been obtained. Finally, power estimation results have been compared with those obtained using XPA when considering the entire system. Indeed, such results served as reference during the comparison. Moreover, we also compare our approach to a classic cumulative approach. It consists in evaluating the power consumption of the entire system based on the sum of the average dynamic powers of each IP. Results are provided in Table \ref{Table_Results_MISO_OFDM_Scenario_label} for the different applications.

\begin{table}[!t]
\caption{Power estimation results of the 4 applications}
\label{Table_Results_MISO_OFDM_Scenario_label}
\centering
\begin{tabular}{|c|c|c|c|c|c|c|}
\hline\hline
 &   XPA  &  Our &   Abs.  &  Cumula- &  Abs.\\
 &   (Ref.)\footnotemark[1] &  \footnotemark[2]&  Error  &   tive \footnotemark[3] &  Error\\ 
 & (mW)  & (mW) & \footnotemark[1]-\footnotemark[2] (\%)  &  (mW) & \footnotemark[1]-\footnotemark[3](\%)\\
\hline\hline
 Appli.& \multirow{2}{*}{118.64}  &  \multirow{2}{*}{122.72} & \multirow{2}{*}{3.44} & \multirow{2}{*}{192.47} & \multirow{2}{*}{62}\\
  1  & & & & & \\
\hline
  Appli. & \multirow{2}{*}{159.01} & \multirow{2}{*}{163.30} &  \multirow{2}{*}{2.7} & \multirow{2}{*}{226.59} & \multirow{2}{*}{42.5}\\
2 & & & & & \\
\hline
  Appli. &  \multirow{2}{*}{195.07} &   \multirow{2}{*}{196.22} &   \multirow{2}{*}{0.59} &  \multirow{2}{*}{266.25} &  \multirow{2}{*}{36.5}\\
3  & & & & & \\
\hline
  Appli. &  \multirow{2}{*}{227.01} &   \multirow{2}{*}{222.11} &   \multirow{2}{*}{2.16}  &  \multirow{2}{*}{294.25} &  \multirow{2}{*}{29.6}\\
4  & & & & &\\
\hline\hline
\end{tabular}
\end{table}

 \begin{figure*}[!t]
 \centering
 \includegraphics[scale=0.55,trim=0cm 0cm 0cm 0cm, clip=true]{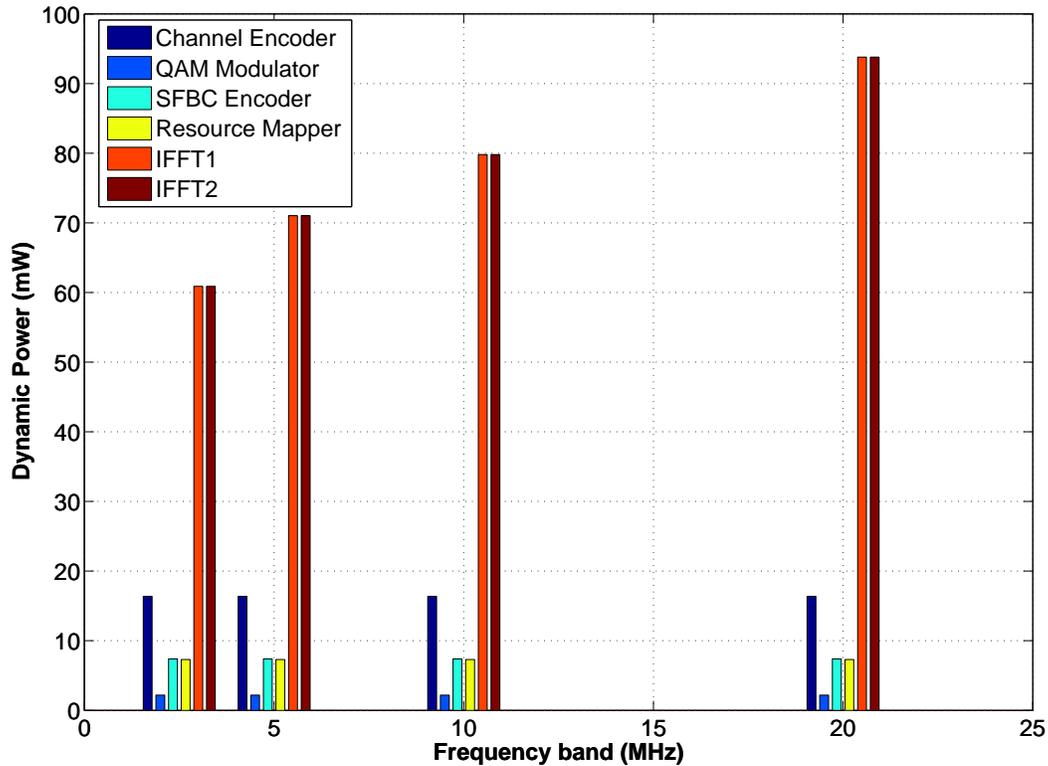}  
 \caption{LTE TX downlink power breakdown in function of frequency band}
 \label{LTE_power_breakdown_label}
 \end{figure*}

In Table \ref{Table_Results_MISO_OFDM_Scenario_label}, it can be noticed that a maximal absolute error lower than 4\% is reached for the four considered applications. Although power estimations are performed at system-level, the obtained accuracy using the proposed approach is really good in comparison to low-level XPA power estimations. Moreover, it can be noted that a very important error is measured using the classic cumulative approach because IP time-activities are not considered. This demonstrates the effectiveness of our approach and the benefits to monitor IP time-activity.

To go further, our tool also permits users to identify the main sources that have an impact on power consumption. As shown in Fig. \ref{LTE_power_breakdown_label}, a power breakdown can be provided to designers for the system under study. From this example, it can be noted that the two IFFTs consume the most significant part of the total power as compared to the other elements in the system. Finally, designers can easily investigate other IPs or hardware options.

\subsection{Speed-up Comparison}

\begin{table}[!t]
\renewcommand{\arraystretch}{1.3}
\caption{Approximated times for power estimation}
\label{Table_Speed_label}
\centering
\begin{tabular}{|c|c|c|c|}
\hline\hline
  &  Proposed &  \multirow{2}{*}{ XPA\footnotemark[1]}  & Speed-up \\
 &  Methodology & &factor \\
\hline\hline
 Appli.1 &  1.25s &  2h25&  x6960\\
\hline
  Appli. 2 &  2s &  6h53 &  x12390\\
\hline
  Appli. 3 &  3.42s & 14h  &  x14736\\
\hline
  Appli. 4 &  6.65s & 27h & x14616\\
\hline\hline
\multicolumn{4}{l}{\footnotemark[1] \footnotesize{Time for timing simulations and XPA analysis} } \\ 
\end{tabular}
\end{table}

\begin{figure*}[t!]
\centering
\begin{subfigure}[b]{0.45\textwidth}
        \includegraphics[scale=0.45,trim=0.5cm 0cm -1cm 0cm, clip=true]{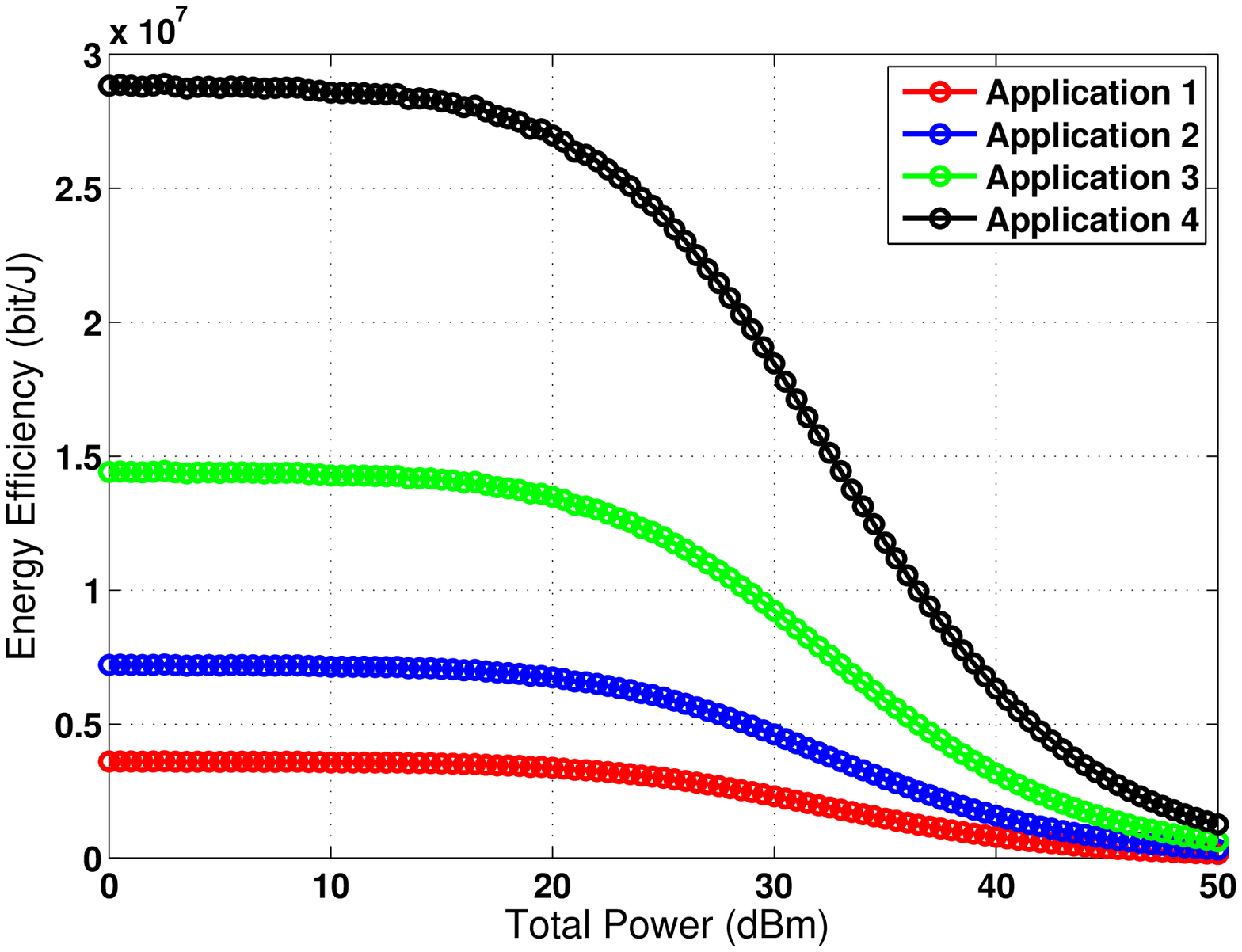} 
        \caption{Average Energy Efficiency (bit/s) versus Total Power (dBm) when $P_{circuit} =0$}
        \label{EE_TX_MISO_OFDM_4applications_less_PC_fig_label}
 \end{subfigure}
    ~ 
    \begin{subfigure}[b]{0.45\textwidth}
        \includegraphics[scale=0.45,trim=0cm 0cm 0cm 0cm, clip=true]{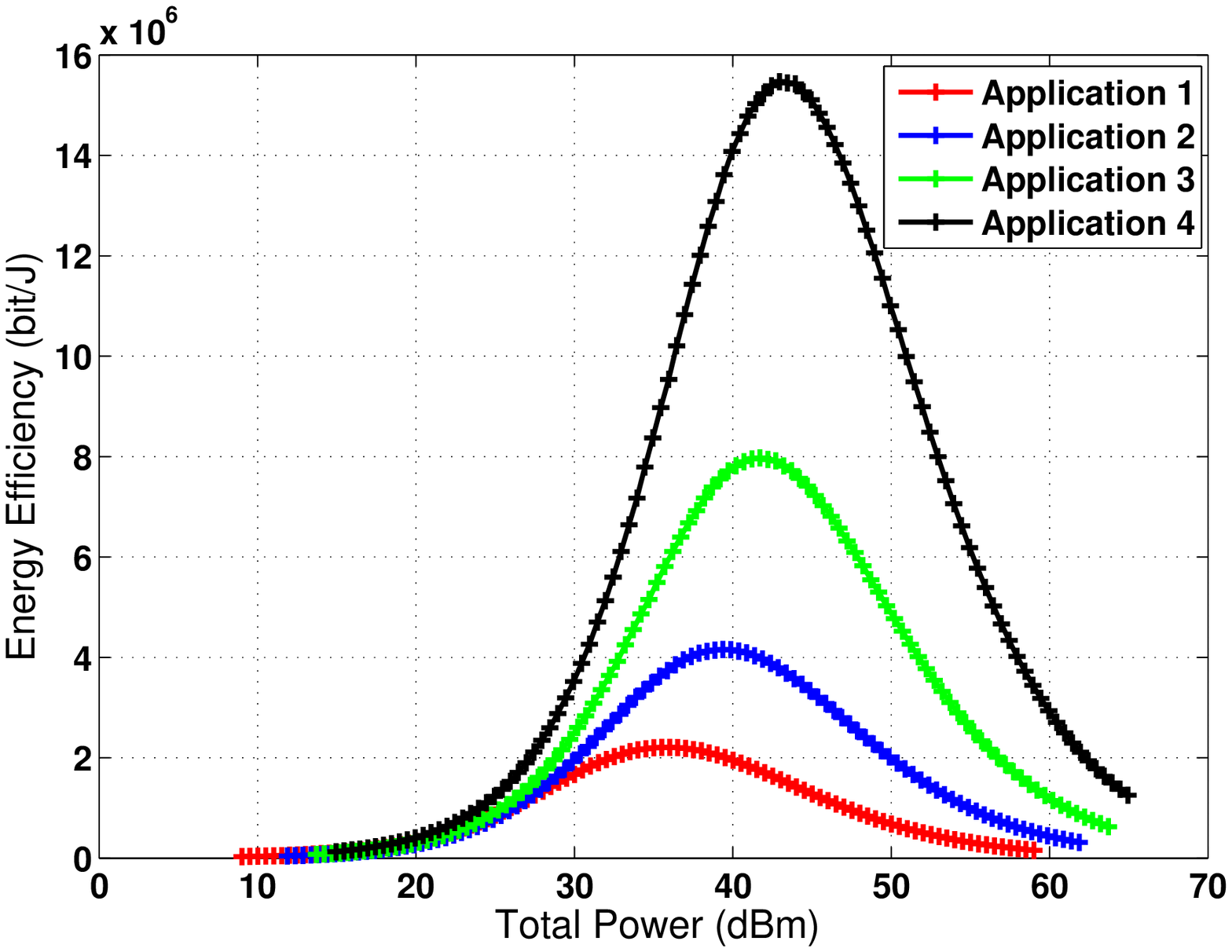} 
        \caption{Average Energy Efficiency (bit/s) versus Total Power (dBm) when $P_{circuit}$ is taken into account}
        \label{EE_TX_MISO_OFDM_4applications_with_PC_fig_label}
    \end{subfigure}
 \end{figure*}

One of the most important benefit of the proposed methodology is the speed-up factor. As indicated in Table \ref{Table_Speed_label}, the usual method for estimating the power consumption of the overall system, which combines timing simulations and XPA analysis, takes several hours. Using our tool, only few seconds (ranging from 1.25s to 6.65s) are required to simulate and to estimate the overall power consumption of the corresponding application. Such differences are generated by the lot of implementation details during low-level simulations of the system. For a same level of accuracy, a speed-up factor of 3-4 order of magnitude is obtained without considering the time that is required during the creation of the library. In most of the cases, designers will directly start their study from the second stage based on the library (that already contains tenth of cores).

The speed-up factor is even more important when many applications have to be tested, especially during design space exploration. In fact, the different steps of the design flow generally have to be rerun from scratch after any parameters modification, such as data quantization, IFFT sizes, etc. Using our approach, only few seconds are required, which corresponds to the compilation of C/C++ files.

\subsection{Energy Efficiency Analysis}

Several domain-specific metrics, such as energy efficiency (EE), can also be easily evaluated during high-level simulations. In fact, this metric reflects the capability of a system to transmit a maximum of data with a minimum of energy. This metric is an efficient way to determine the best power-performance trade-off. Using our approach, additional specific metrics of the wireless communication domain can be obtained such as the Bit-Error Rate (BER), etc. 
 
EE is usually evaluated as follows:
\begin{equation} \label{eq_EE_MISO_label}
EE (bit/J) = \frac{C}{P_{Total}}= \frac{W.E[log2(1+ \frac{|h|^2 P_{t} PL}{N0 W Nt})]}{P_{Total}} \\
\end{equation}
with $C$ the average capacity for a MISO configuration (bit/s), $W$ the frequency bandwidth (Hz), $h$ the fading coefficients of the channel, $N0$ the noise spectral density (dBm/Hz), $Nt$ the number of transmit antennas i.e. 2, $PL$ the path loss (dB), $P_{Total}$ the average total power (W) that is consumed by the system and where:
 \begin{equation}
P_{Total}= P_{t} + P_{circuit} \\
\end{equation} 
with $P_{t}$ the power allocated for data transmission and $P_{circuit}$ the average dynamic power that is consumed by the circuit (i.e. the base-band processing in our study).
From Eq. \ref{eq_EE_MISO_label}, we assume that $PL/N0.W =1$ during this study. This assumption is fully-compliant with a current small-cell (micro or pico) environment in LTE standard (for a frequency of 2.6GHz, $N0 = -174dBm$ and the LTE frequency bands). 

EE has been computed and is represented in Figures \ref{EE_TX_MISO_OFDM_4applications_less_PC_fig_label} and \ref{EE_TX_MISO_OFDM_4applications_with_PC_fig_label}. Two cases have been considered. First, EE has been evaluated when the power consumption of the circuit was not considered, i.e. $P_{circuit}= 0$. Only $P_{t}$, the power allocated for data transmission, was thus taken into account during the EE computations. Secondly, the power consumption related to the circuit was considered during EE evaluations. 
It can be noticed that the power consumption of the circuit has a significant impact on the EE. 
Moreover, it is interesting to note that application 4 has the most important power consumption and is also the most energy efficient application. This is because this application has the largest capacity and frequency bandwidth. Note that the energy efficiency of the four applications are really close to each other at low transmit power. Through this example, we demonstrate that all sources of power consumption have to be taken into account in order to provide realistic results. 
 
\section{Conclusion and Future works} \label{sect_conclu}

In this paper, a system-level power estimation methodology for FPGA-based hardware designs has been presented. Fast and accurate power estimations of a FPGA-based wireless communication transmitter in LTE context have been described. 
The innovative concept of scenario was also introduced. This concept is different from the examples found in the literature. It aims at helping designers to efficiently compare several design choices and to observe the impact of a specific parameter on power consumption. Another major contribution was the monitoring of IP time-activities that enable to refine power estimations. We also provide to designers a dedicated library of hardware IPs with their corresponding high-level models. All these contributions enable designers to perform an efficient and fast design space exploration.
Energy efficiency of several wireless communication transmitters has also been evaluated. The results highlight the impact of the base-band power consumption and enable designers to choose the most energy efficient system for a given power consumption. To design future wireless communication systems, such power consumption information has to be taken into account by designers in order to satisfy both power and performance requirements. This is of particular importance when low transmission powers are considered.

As future works, the proposed approach will be used to compare several base-band processing schemes of various wireless communication systems. Power and performance metrics as energy efficiency will be evaluated. Moreover, additional wireless communication elements will be included such as the power amplifier or the RF stages. 
The main limitation of the methodology is currently being investigated. Indeed, we aim at generalizing the power estimation values that have been obtained on a dedicated FPGA to other FPGA families. We will also refine power estimations using real measurements as compared to the current low-level power estimations.


\section*{Acknowledgement}

The authors would like to thank Orange Labs for their financial support.

\bibliographystyle{abbrv}
\bibliography{biblio_HIP3ES}
%



\end{document}